\documentclass[reprint, superscriptaddress, preprintnumbers, nofootinbib, amsmath,amssymb, aps]{revtex4-2}

\usepackage{graphicx}
\usepackage{bm}
\usepackage{subfigure}
 \usepackage{mathrsfs}
\usepackage{hyperref}
\usepackage{booktabs}
\usepackage{rotating}
\usepackage{fontawesome}

\makeatletter
\newcommand{\github}[1]{%
   \href{#1}{\faGithubSquare}%
}
\makeatother

\usepackage{scalerel}
\usepackage{tikz}
\usetikzlibrary{svg.path}

\definecolor{orcidlogocol}{HTML}{A6CE39}
\tikzset{
  orcidlogo/.pic={
    \fill[orcidlogocol] svg{M256,128c0,70.7-57.3,128-128,128C57.3,256,0,198.7,0,128C0,57.3,57.3,0,128,0C198.7,0,256,57.3,256,128z};
    \fill[white] svg{M86.3,186.2H70.9V79.1h15.4v48.4V186.2z}
                 svg{M108.9,79.1h41.6c39.6,0,57,28.3,57,53.6c0,27.5-21.5,53.6-56.8,53.6h-41.8V79.1z M124.3,172.4h24.5c34.9,0,42.9-26.5,42.9-39.7c0-21.5-13.7-39.7-43.7-39.7h-23.7V172.4z}
                 svg{M88.7,56.8c0,5.5-4.5,10.1-10.1,10.1c-5.6,0-10.1-4.6-10.1-10.1c0-5.6,4.5-10.1,10.1-10.1C84.2,46.7,88.7,51.3,88.7,56.8z};
  }
}

\newcommand\orcidicon[1]{\href{https://orcid.org/#1}{\mbox{\scalerel*{
\begin{tikzpicture}[yscale=-1,transform shape]
\pic{orcidlogo};
\end{tikzpicture}
}{|}}}}

\usepackage[]{xcolor}
\usepackage{hyperref}
\hypersetup{
    colorlinks = true,
    citecolor = {magenta},
    linkcolor = {blue},
    urlcolor = {blue}		
}
\usepackage{mathtools}
\usepackage{aas_macro}
\usepackage{xspace}

\def\bi#1{\hbox{\boldmath{$#1$}}}

\newcommand\gpch{$h^{-1} {\rm Gpc}$\xspace}
\newcommand\mpch{$h^{-1} {\rm Mpc}$\xspace}
\newcommand\msunh{$h^{-1} M_\odot$\xspace}
\newcommand\hmpc{$h {\rm Mpc}^{-1}$\xspace}

\begin{document}

\title{The HalfDome Multi-Survey Cosmological Simulations: N-body Simulations}

\author{Adrian E.~Bayer \orcidicon{0000-0002-3568-3900}}
\email{abayer@princeton.edu}
\thanks{Co-first author}
\affiliation{Department of Astrophysical Sciences, Peyton Hall, Princeton University, Princeton, NJ 08544, USA}
\affiliation{
Center for Computational Astrophysics, Flatiron Institute, 162 5th Avenue, New York NY 10010, USA
}

\author{Yici Zhong \orcidicon{0000-0003-0805-8234}}
\email{yici.zhong@phys.s.u-tokyo.ac.jp}
\thanks{Co-first author}
\affiliation{
Department of Physics, School of Science, The University of Tokyo, 7-3-1 Hongo, Bunkyo-ku, Tokyo 113-0033, Japan
}

\author{Zack Li  \orcidicon{0000-0002-0309-9750}}
\affiliation{Lawrence Berkeley National Laboratory,  1 Cyclotron Road, Berkeley, CA 94720, USA}
\affiliation{
 Berkeley Center for Cosmological Physics, University of California,
Berkeley,  CA 94720, USA
}%
\affiliation{Canadian Institute for Theoretical Astrophysics, University of Toronto, Toronto, ON, M5S 3H8 Canada}

\author{Joseph DeRose}
\affiliation{Lawrence Berkeley National Laboratory,  1 Cyclotron Road, Berkeley, CA 94720, USA}

\author{Yu Feng}
\affiliation{
Berkeley Center for Cosmological Physics, University of California,
Berkeley,  CA 94720, USA
}%

\author{Jia Liu \orcidicon{0000-0001-8219-1995}}
\affiliation{
Center for Data-Driven Discovery, Kavli IPMU (WPI), UTIAS, The University of Tokyo, Kashiwa, Chiba 277-8583, Japan
}
\affiliation{
Kavli IPMU (WPI), UTIAS, The University of Tokyo, Kashiwa, Chiba 277-8583, Japan
}

\date{\today}

\begin{abstract}

Upcoming cosmological surveys have the potential to reach groundbreaking discoveries on multiple fronts, including the neutrino mass, dark energy, and inflation. Most of the key science goals require the joint analysis of datasets from multiple surveys to break parameter degeneracies and calibrate systematics. To realize such analyses, a large set of mock simulations that realistically model correlated observables is required. In this paper we present the N-body component of the HalfDome cosmological simulations, designed for the joint analysis of Stage-IV cosmological surveys, such as Rubin LSST, Euclid, SPHEREx, Roman, DESI, PFS, Simons Observatory, CMB-S4, and LiteBIRD. 
Our 300TB initial data release includes full-sky lightcones and halo catalogs between $z$=0--4 for 11 fixed cosmology realizations, as well as an additional run with local primordial non-Gaussianity ($f_{\rm NL}$=20). 
The simulations evolve $6144^3$ particles in a 3.75$\,h^{-1} {\rm Gpc}$ box, reaching a minimum halo mass of $\sim 6 \times 10^{12}\,h^{-1} M_\odot$ and maximum scale of $k \sim 1\,h{\rm Mpc}^{-1}$. Our data is publicly available:
instructions to access the data and plans for future data releases can be found at \url{https://halfdomesims.github.io}.

\end{abstract}

\maketitle

\section{Introduction}

\label{sec:intro}

Ongoing and upcoming cosmological surveys will measure the growth of structure and geometry of our universe at unprecedented precision, with the ambition to probe physics beyond the standard model on multiple fronts. While individual surveys can provide unique and valuable information, combined analysis of multiple datasets, such as weak gravitational lensing, galaxy clustering, and the cosmic microwave background (CMB), is required to answer the most pressing questions in cosmology. For example, to measure the neutrino mass, no single cosmological survey can probe the neutrino mass sum at the $5\sigma$ level if it is at the minimum possible value of 0.06~eV~\cite{Font-Ribera2014,Schmittfull_2018, Mishra-Sharma_2018, Bayer_2022_fake, Qu_2023}. The motivations to combine data from multiple surveys include: (1) probing otherwise undetectable signals, (2) increasing the sensitivity to time- and scale-dependent deviations from $\Lambda$CDM, (3) increasing the overall signal-to-noise, (4) breaking the degeneracies between cosmological and astrophysical parameters, and (5) mitigating systematics. To realize the full potential of joint analyses, we need to overcome not only the usual challenges faced by individual surveys but also unique challenges related to the correlated signals and systematics between them. Most of these correlations are complex and nonlinear, and hence realistic mock simulations that jointly model observables from many experiments are critical in enabling modern cosmological analyses with multiple datasets. 

Joint analyses place greater demands on simulations compared to those tailored to individual surveys~\cite{Shirasaki2019:HSCsims,Korytov2019:DC2, DeRose2022:DESsims, Grove2022:DESIsim, To2023}. First, Stage-IV surveys\footnote{The concept of Stage-IV surveys was introduced in the Dark Energy Task Force report~\cite{DETF}, referring to the cosmological experiments starting in the 2020s.} will cover a large portion of the sky and probe the matter distribution between $0<z\lesssim4$, thus we require the simulations to cover a very large volume. Second, to implement realistic galaxy formation and CMB foreground physics, the simulations need to reach sufficient resolution for all surveys in consideration. Third, as the surveys overlap in sky regions and redshifts, cosmological signals and systematics are correlated (e.g. between CMB foregrounds and the galaxy distribution), thus requiring different astrophysical observables in the simulations to be correlated with each other. Finally, to minimize the variance of measured statistics and to compute covariances, many realizations of such simulations must be produced. 
Previous joint simulations have partially addressed these requirements~\citep{Sehgal_2010,Han:2021unz,McCarthy2017,McCarthy2018, Takahashi2017,Liu2018,Derose2019:Buzzard,Stein:2020its,Omori:2022uox}.

Here we present the HalfDome cosmological simulations -- a set of simulations tailored for the joint analysis of Stage-IV surveys, such as the Vera Rubin Observatory LSST~\cite{LSST}, Euclid~\cite{Euclid}, SPHEREx~\cite{SPHEREx:2018xfm}, Roman Space Telescope~\cite{RST}, DESI~\cite{DESI}, PFS~\cite{PFS}, Simons Observatory~\cite{SimonsObservatory:2018koc}, CMB-S4~\cite{CMB-S4:2016ple}, and LiteBIRD~\cite{LiteBIRD:2022cnt}. For our initial data release, our choice of resolution, sky coverage, observables, and the number of simulations are designed to balance the different requirements of large-scale structure (LSS) and CMB surveys, within the current computational limitations. In particular we release 11 simulations at a fixed cosmology to enable uncertainty quantification in survey analysis pipeline testing. To allow continuous future upgrades, our simulation pipeline is designed with distinct modules, making it possible to flexibly adopt future advancements in observations, numerical methods, and theories. Thus, while the initial data release considers a particular cosmology with a particular N-body simulation and a particular halo finder, we can easily run more simulations with different settings and plug these into the same pipeline to produce CMB and LSS data products. Table \ref{table:sims} shows a comparison of the HalfDome specifications with other publicly available simulations for joint analyses. 

\begin{table*}
\begin{center}
\renewcommand{\arraystretch}{1.25}
{\begin{tabular}{ l | r | r | r | r | r | r | r | r}
\toprule
& Sehgal+ & BAHAMAS & Takahashi+ & MassiveNuS & Buzzard & Websky  & Agora  & HalfDome \\[2pt]
& \cite{Sehgal_2010,Han:2021unz} & \cite{McCarthy2017,McCarthy2018} & \cite{Takahashi2017} & \cite{Liu2018}& \cite{Derose2019:Buzzard} & \cite{Stein:2020its} & \cite{Omori:2022uox} & (This work) \\[2pt]
\hline
$L_{\rm box}$~(\gpch) 	    & 1 & 0.4  & 0.45--6.3 & 0.512 & 1.05--4.0 & 7.7 & 1 & 3.75 \\ 
$N_c$              & $1024^3$ & $2\times1024^3$& $2048^3$ ($z$=1) & $1024^3$& $2048^3$ ($z$=1) & $6144^3$ & $3840^3$ & $6144^3$ \\ 
$M_h^{\rm min}$ (\msunh)   & $10^{13}$& $5\times 10^{11}$ & $10^{13}$ ($z$=1)& $10^{10}$ & $10^{13}$ ($z$=1) & $1.2\times 10^{13}$ & $1.5\times 10^{9}$ & $6 \times 10^{12}$ \\ 
$N_{\rm sim}$             & 1 (500) & 11 & 108 & 101 & 30  & 1 & 1 & 11 (+1 $f_{\rm NL}$) \\ 
\toprule
\end{tabular}}
\caption{\label{table:sims} 
Comparison of HalfDome simulations to existing public simulations with correlated CMB and LSS mocks, in terms of the box size $L_{\rm box}$, number of particles $N_c$, minimum halo mass $M_h^{\rm min}$, and number of simulations $N_{\rm sim}$. The Takahashi and Buzzard simulations have nested boxes with varying resolutions, therefore we only quote $N_c$ and $M_h^{\rm min}$ at $z$=1. We also note that \cite{Han:2021unz} used machine learning to make 500 realizations out of the 1 realization from \cite{Sehgal_2010} excluding the halo catalog.}
\end{center}
\end{table*}

In this first paper of the series, we describe the N-body component of the HalfDome simulations. We provide detailed descriptions of the codes we used, simulation setup, data products, data structures, run time, and validation results. To provide a playground for upcoming cosmological surveys, we make an initial data release of 300TB, including particle lightcones, halo catalogs, and mass and velocity sheets. Our forthcoming papers will present realistic mocks that model multi-wavelength observables such as galaxies, weak gravitational lensing, and CMB foregrounds, as well as simulations with different cosmologies. Our data is publicly available:
instructions to access the data and plans for future data releases can be found at \url{https://halfdomesims.github.io}.

This paper is organized as follows. We introduce our method in Sec.~\ref{sec:method}, including the N-body code \texttt{FastPM} and simulation outputs. In Sec.~\ref{sec:result} we show our results and validation against theory, in particular we dedicate Sec. \ref{sec:fnl} to discuss an application of our simulations to primordial non-Gaussianity. Finally, we summarize our work and discuss future prospects in Sec.~\ref{sec:conclusion}. 

\section{Method}
\label{sec:method}

In this section, we describe the N-body simulation code \texttt{FastPM}\footnote{\url{https://github.com/fastpm/fastpm}}~\cite{Feng2016, Bayer_2021_fastpm} employed in this work. We also describe the methods used to downsample particles, find halos, generate lightcones, and compute mass and velocity sheets. Details of the HalfDome simulations are summarized in Table~\ref{Tab:sims}.

\subsection{\texttt{FastPM} N-Body Simulations}

\texttt{FastPM} is a particle mesh simulation code for LSS. It achieves faster convergence than standard particle mesh simulations by modifying the kick and drift factors to enforce the Zel’dovich approximation in the time evolution. Furthermore, it employs a grid-based Fourier kernel \cite{Hockney_1988} (instead of the naïve choice of $\nabla = ik$ and $\nabla^2 = -k^2$) to avoid the need for corrections during mass assignment. This enables the generation of simulations with a much shorter computation time than a full tree-PM approach but with comparable accuracy. As such, \texttt{FastPM} has been employed for analyzing multiple recent cosmological datasets. For example the DESI collaboration \cite{ 2013arXiv1308.0847L} actively uses it to generate mock galaxy catalogs, and has performed thorough tests of its accuracy compared to full N-body simulations showing excellent agreement into the nonlinear regime \cite{Ding_2022, Variu_2023}. Additionally, its massive neutrino implementation \cite{Bayer_2021_fastpm} has also been used to analyze voids in the BOSS DR12 data \cite{2023arXiv230707555T}. Its combination of speed and accuracy has facilitated hierarchical Bayesian reconstruction of the initial cosmic field \cite{Modi_2018, Bayer_2023_vel} and field-level inference \cite{MCLMC} using GPU adaptations of the code \cite{Modi_2021, li2022pmwd}. It has also been used to create a suite of simulations to measure super-sample covariance \cite{Bayer_2023_ssc}.

The HalfDome initial data release runs have a box length of 3.75~\gpch with $N_c$=6144$^3$ cold dark matter particles and a particle mass of $1.95\times 10^{10}~$\msunh. We use cosmological parameters
$h$=0.6774, 
$\Omega_m$=0.3089, 
$\Omega_b$=0.0486,
$\sigma_8$=0.8159,
$n_s$=0.9667,
$M_\nu$=0, matching IllustrisTNG \cite{2019ComAC...6....2N}, and initialize the simulation with the linear matter power spectrum at $z$=0 computed with \texttt{CLASS}~\cite{classy}.
The simulations start at $z$=9 and take 60 time steps that are linearly spaced in $a$ until $z$=0. The force mesh resolution is taken to be twice that of the particle grid, i.e.~$12288^3$. 
Prior to generating the full-resolution runs, we ran several lower-resolution simulations for code validation.  See Table~\ref{Tab:sims} for a summary for both the full- and lower-resolution runs. 

\begin{table*}
\centering
\begin{tabular}{c|c|c|c|c|c|c|c|c|c|c|r|r|r}
\toprule
\multicolumn{1}{c}{} & \multicolumn{1}{c}{} & \multicolumn{3}{c}{\textbf{Simulation setup}} & \multicolumn{6}{c}{\textbf{Run time}} & \multicolumn{3}{c}{\textbf{Data volume}} \\
\cmidrule(rl){3-5} \cmidrule(rl){6-11} \cmidrule(rl){12-14}
Run & \textbf{$N_{\rm sim}$} & {$N_c$} & {$L_{\rm box}$ [$h^{-1}{\rm Gpc}$]} & $N_{\rm side}$ & $N_{\rm node}$ & {Hour} & IC [$\%$] & PM [$\%$] & FoF [$\%$] & IO [$\%$] & Halos& Particles & Sheets\\
\midrule

full res & 11 (+1) & $6144^3$ & $3.75$ & $8192$ & $2048$ & $4.38$ & $2.89$ & $72.40$ & $14.48$ & $10.24$  & $362$G & $17$T & $1.4$T\\

1/2 res & 1 & $4096^3$ & $5$ & $8192$ & $512$ & $(1.95)$ & $3.26 $ & $74.17 $ & $10.84 $ & $11.73$  & $29$G & $7.7$T & $824$G\\

1/4 res & 1 & $2048^3$ & $5$ & $8192$ & $64$ & $2.42$ & $3.72 $ & $81.61 $ & $11.92 $ & $2.76 $  & $1.4$G & $1.2$T & $376$G\\

1/8 res & 1 & $1024^3$ & $5$ & $2048$ & $8$ & $2.32$ & $3.08 $ & $77.22 $ & $18.81 $ & $0.89$  & $21$M & $149$G & $33$G\\

1/16 res & 1 & $512^3$ & $5$ & $2048$ & $1$ & $3.01$ & $1.73 $ & $65.86 $ & $32.21$ & $0.21$  & $176$K & $19$G & $8.9$G\\
\bottomrule
\end{tabular}
\caption{\label{Tab:sims}Summary of the HalfDome simulation N-body runs, including the number of simulations $N_{\rm sim}$, particle number $N_c$, box size $L_{\rm box}$, HEALPix map resolution $N_{\rm side}$, number of computational nodes $N_{\rm node}$, run time, and data volume (per simulation). The $1/2$ resolution run was performed on NERSC Cori KNL nodes, while other simulations were performed on Stampede2 KNL nodes, hence the timing of the 1/2 resolution run is in brackets to avoid confusion in timing scaling. 
At the full resolution, in addition to the 11 fixed-cosmology runs, we provide an additional run with primordial non-Gaussianity $f_{\rm NL}$=20.
}
\end{table*}

\subsection{Lightcones}
 \label{sec:lightcone}

We generate a full-sky lightcone for each simulation.
We tile the 3.75~\gpch box approximately 2.6 times per dimension to cover the volume between $z$=0--4.
The lightcone is populated with particles on-the-fly at the appropriate redshift, interpolating between N-body time steps.

\begin{figure}
   \centering
    \includegraphics[width=1.\linewidth]{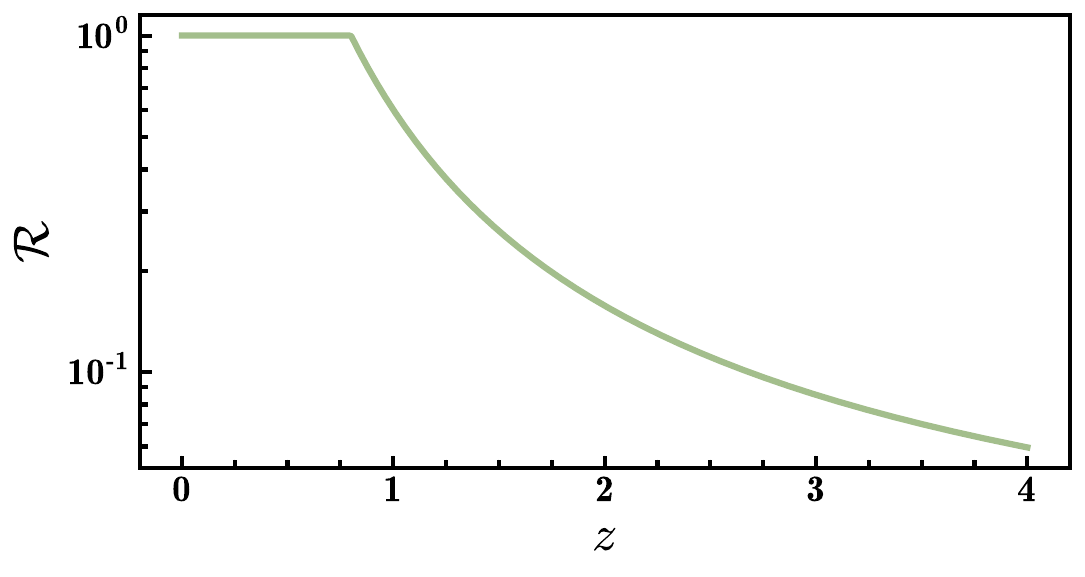}
    \caption{Downsampling rate (Eq.~\ref{eq:downstample}) for our full-resolution runs, with $N_c$=$6144^3$, $L_{\rm box}$=3.75~\gpch, and $\ell_{\rm max}$=10,000.}
    \label{fig:s_r}
\end{figure}

To reduce the data footprint, we downsample the lightcone particles with a redshift-dependent rate, 
\begin{align} \label{eq:downstample}
    \mathcal{R}(z) =& \min\left[1,\left(\frac{\theta_{\rm original}(z)}{\theta_{\rm min}} \right)^3 \right],
\end{align}
where $\theta_{\rm original}$ is the average angular separation of particles in the lightcone and $\theta_{\rm min}$ is the minimum angular separation considered. Mathematically,
\begin{align}
    \theta_{\rm original}(z)=&\frac{L}{N^{1/3}_c\chi(z)},\\
    \theta_{\rm min} =& \frac{\pi}{\ell_{\rm max}},
\end{align}
where $L_{\rm box}$ is the box length, $N_c^{1/3}$ is the number of particles per side (e.g. 6144 for our full resolution run), $\chi(z)$ is the comoving distance at redshift $z$, and $\ell_{\rm max}$ is the maximum angular wavenumber we sample. 
We choose $\ell_{\rm max}$=10,000 to roughly double the maximum angular wavenumber expected to be used in Stage-IV surveys\footnote{For example CMB-S4 will be limited by planned sensitivity and instrumental resolution to $\ell_{\rm max}\sim$ 5,000 \cite{CMB-S4:2016ple}.}. The downsampling rate $\mathcal{R}(z)$ for our full resolution runs is shown in Fig.~\ref{fig:s_r}.

\subsection{Halos}
To find halos we use the Relaxed-Friends-of-Friends (RFoF) algorithm~\cite{2020JCAP...04..002D}. RFoF works much like FoF, but with a linking length that varies with redshift and halo mass. In particular, RFoF increases the linking length for halos with fewer particles to enable finding low-mass halos, with as low as 20 particles, that would otherwise be missed with the traditional FoF halo finder. However, after using the default RFoF hyperparameters we found it contained numerical artifacts for halos below 320 particles, corresponding to a mass of $\sim 6\times10^{12}$\msunh, and therefore deem any halos below this mass to be unphysical.
Our motivation for using RFoF is twofold: first, it is built-in to \texttt{FastPM} so that halos can be found on-the-fly in a computationally efficient manner, and second, it can find halos that traditional Friend-of-Friends (FoF) cannot. In future data releases we plan to implement different halo finders, such as ROCKSTAR \cite{Behroozi:2011ju}.

While RFoF returns the halo mass based on the number of particles in a halo, it is known that the mass function of FoF halos does not follow the commonly-used mass function of \cite{Tinker2008}~(Tinker08). We thus also provide an alternative definition of halo mass by calibrating the halo mass to $M_{200m}$ by abundance matching with Tinker08. To do so, we compute a smooth function of RFoF mass and redshift, $M_{200m}(M_{\rm RFoF}, z)$. We compute this smooth function by extracting narrow redshift slices of $\Delta z = 0.05$ from the lightcone and generating a matching realization of the Tinker08 mass function at that redshift with the same volume. We abundance match the redshift slice to the desired mass function and then fit a smooth function $M_{200m}(M_{\rm RFoF})$ at that redshift. In practice, we compute the mean $M_{200m}$ for bins of RFoF masses, with steps of $\Delta \log_{10} M_{\rm RFoF}/M_{\odot} = 0.05$ and then apply a Savitzky-Golay filter to reduce noisy fluctuations at high masses. We repeat this process at each redshift and then perform a bivariate spline interpolation to generate $M_{200m}(M_{\rm RFoF}, z)$. The high mass tail of the mass function has relatively few halos, so we perform a polynomial fit on the high mass tail with Poisson error weighting to extrapolate to higher masses. 
The mass definition is accurate for halos with more than 320 particles, corresponding to a minimum mass of $\sim 6 \times 10^{12}$~\msunh. A similar approach of recalibrating mass definitions was performed for the Euclid Flagship simulation \cite{Euclid:2024few}. 

\begin{figure*}
    \centering
    \includegraphics[width=0.9\textwidth]{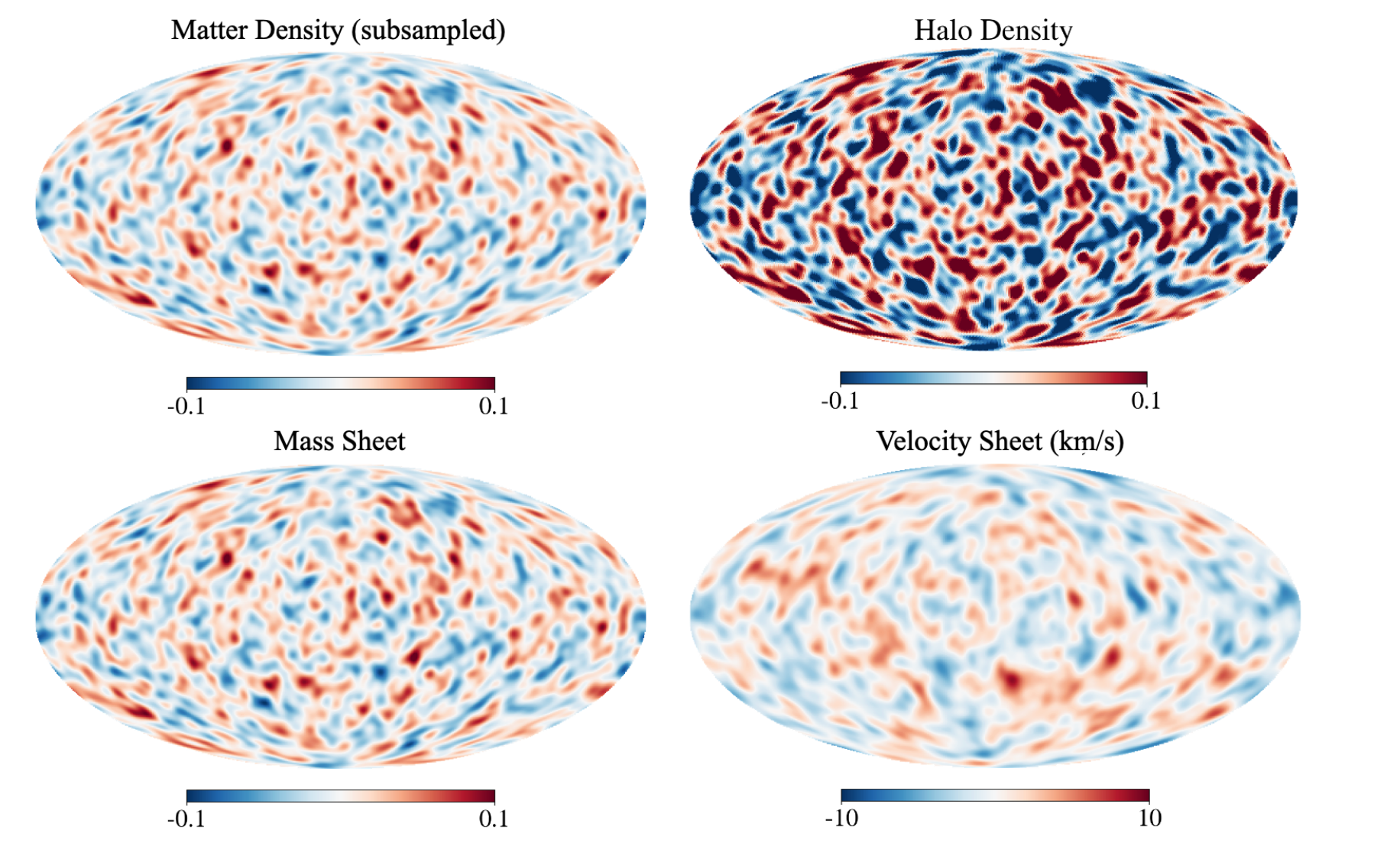}
    \caption{Sample HalfDome data products at $z$=1 from one of the full-resolution runs. {\bf Top left}: matter overdensity $\delta = (\rho-\bar{\rho})/\bar{\rho}$ obtained using downsampled particles from the lightcone. 
    {\bf Top right}: halo overdensity $\delta_h = (n_h-\bar{n}_h)/\bar{n}_h$ with a minimum halo mass $6 \times 10^{12}$~\msunh. 
    {\bf Bottom left}: mass sheet (without particle downsampling) which is the projected matter density in a redshift slice. 
    {\bf Bottom right}: velocity sheet which shows the mean radial velocities of particles within a redshift slice. All maps have a thickness $\Delta a$ = 0.01.
    }
    \label{fig:maps}
\end{figure*}

\subsection{Mass and velocity sheets}
\label{sec:sheets}
We generate mass and velocity sheets using the full-resolution lightcones (Sec.~\ref{sec:lightcone}). Mass sheets are two-dimensional (2D) shells of projected matter density. Velocity sheets are 2D shells of mean radial velocities of particles within the redshift slice. Both mass and velocity sheets are produced in HEALPix\footnote{\url{https://healpix.sourceforge.io/}} \cite{Gorski:2004by} format with $N_{\rm side}=8192$ for the full-resolution runs. 80 mass and velocity sheets are generated between $a$=0.2--1 ($z$=4--0), with $\Delta a$ = 0.01. These will be useful for the construction of projected quantities, such as weak lensing or CMB foreground maps.


\section{Results}
\label{sec:result}

Fig.~\ref{fig:maps} shows sample maps of the matter density, halo density, mass sheet, and velocity sheet at redshift $z$=1, with thickness $\Delta a$=0.01 from one of the full resolution runs. In this section, we show validation test results for the matter field (Sec.~\ref{sec:pk}) and halo field (Sec.~\ref{sec:halo}). We also present one additional run with primordial non-Gaussianity  (Sec.~\ref{sec:fnl}).  

\subsection{Matter field} 
\label{sec:pk}
The power spectrum $P(k)$ is defined as the Fourier transform of the two-point correlation function $\xi\left(r\right)$,
\begin{align}
P(k) & = \int d^3{\bi{r}} ~ \xi(r) e^{i \bi{k} \cdot \bi{r}},\\
\xi\left(r\right) & =  \xi\left(|\bi{r}|\right) = \left\langle\delta\left(\bi{x}\right) \delta\left(\bi{x}+\bi{r}\right)\right\rangle,\\
\delta({\bi{x}}) & =\frac{\rho({\bi{x}})-\bar{\rho}}{\bar{\rho}}, 
\end{align}
where $r$ is the distance, $\rho(\bi{x})$ is the matter density, and $\bar{\rho}$ is the mean density. To compute the power spectrum of the simulations, the particles are painted to a grid using the Cloud-in-Cell interpolation to generate the matter overdensity field $\delta$. 



The angular power spectrum is given by
\begin{align}
\label{eq:cldd}
C_\ell=&\int d\chi_1 d\chi_2F(\chi_1)F(\chi_2) \int\frac{2k^2dk}{\pi}j_\ell(k\chi_1) j_\ell(k\chi_2) P(k),
\end{align}
where $\chi$ is the comoving distance, $j_\ell$ are the spherical Bessel functions, and $F(\chi)$ is the projection kernel,
\begin{equation}
    F(\chi(z)) = \frac{H(z)}{c} W(z) D(z),
\end{equation}
where $\chi=\chi(z)$ is the comoving distance to redshift $z$, $H(z)$ is the Hubble function, $W(z)$ is the window function (which we take as a top hat for plots in this paper), and $D(z)$ is the linear growth function.

\begin{figure*}
    \centering
    \includegraphics[width=1.0\textwidth]{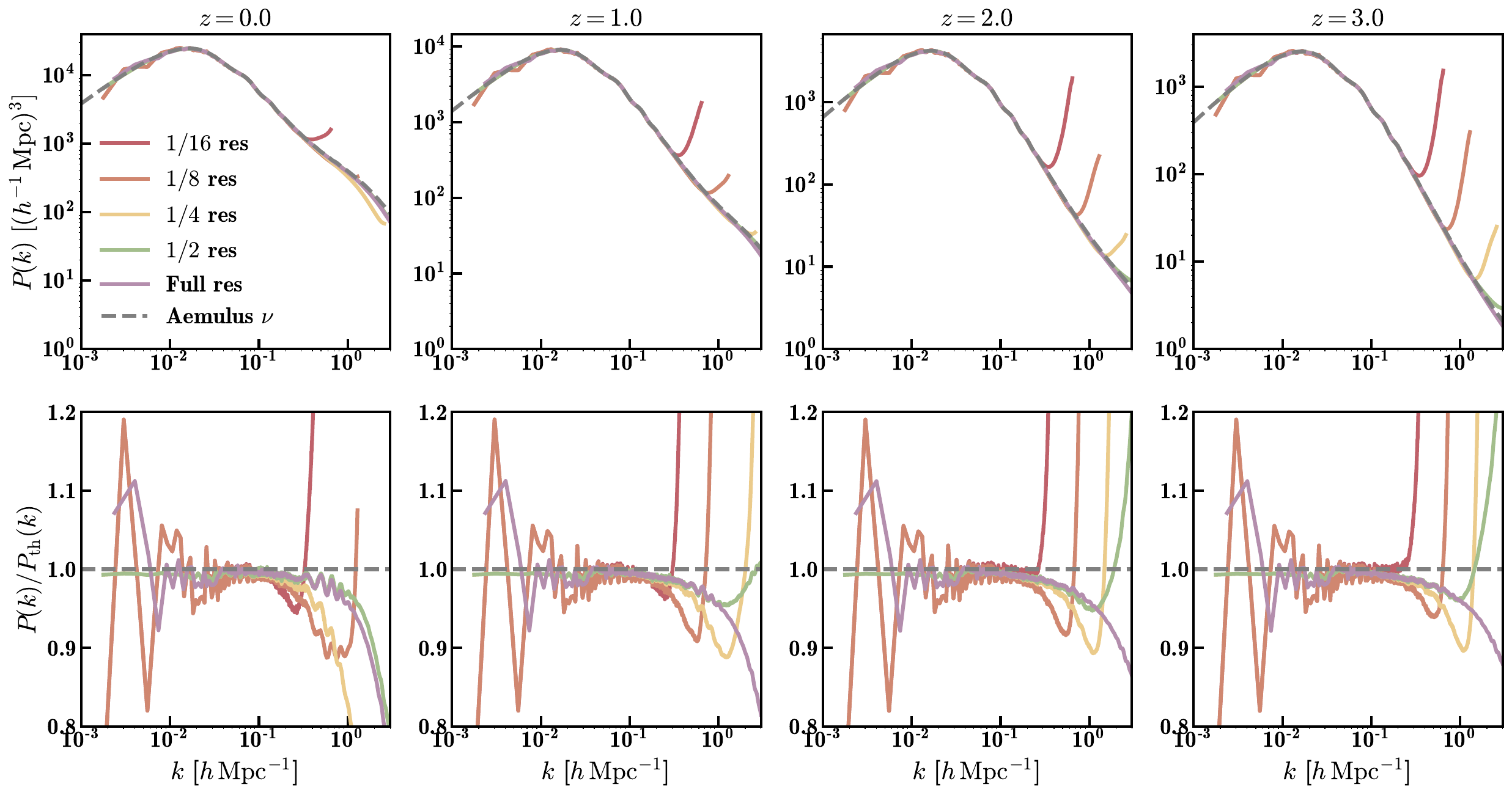}
    \caption{{\bf Top panels}: matter power spectra of HalfDome simulations of different resolutions and the theoretical prediction from Aemulus $\nu$, at $z$=0, 1, 2, 3. {\bf Bottom panels}: the ratios between the simulations and Aemulus $\nu$.
    }
    \label{fig:Pk}
\end{figure*}

In Fig.~\ref{fig:Pk}, we show the matter power spectra computed from our simulations of different resolutions, as well as the theoretical prediction from Aemulus $\nu$~\cite{Aemulus}
in the top panels, for $z$=0, 1, 2, 3.  In the bottom panels, we show the ratios of the simulated power spectra relative to Aemulus $\nu$. Our full-resolution simulations agree with Aemulus $\nu$ within 4\% up to $k$$\sim$1~\hmpc. 

\begin{figure*}
    \centering
    \includegraphics[width=1.0\textwidth]{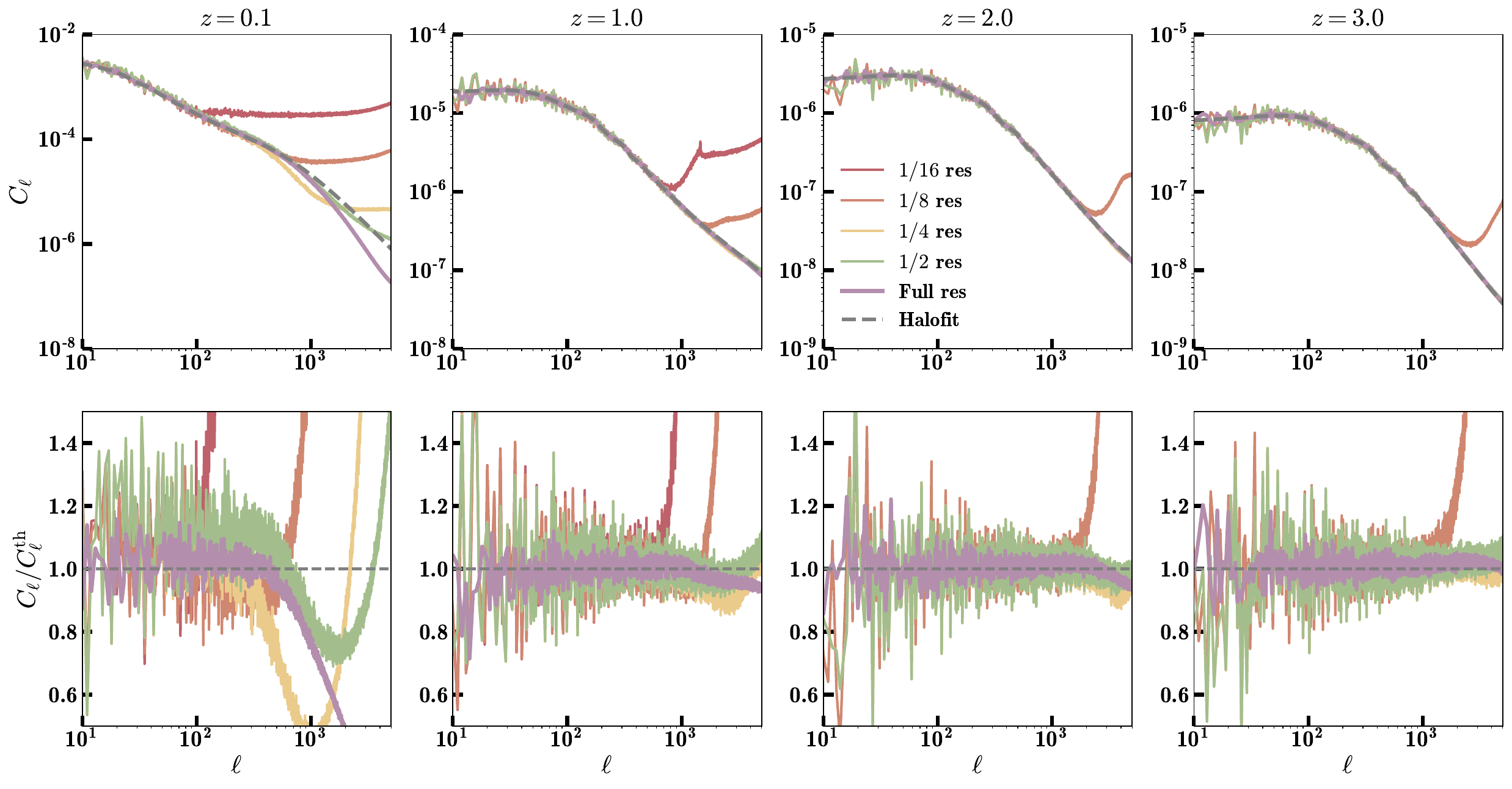}
    \caption{
    {\bf Top panels}: angular power spectra of  HalfDome simulations of different resolutions and the theoretical prediction from Halofit, at $z$=0, 1, 2, 3. {\bf Bottom panels}: the ratios between the simulations and Halofit.
    }
    \label{fig:cldd}
\end{figure*}

Similarly, in Fig.~\ref{fig:cldd} we show the angular matter power spectra $C_\ell$ at different resolutions. We again compare it with the theoretical prediction from Halofit. Our full-resolution simulations agree with Halofit at the 10\% level at least to $\ell \sim 5$,000 for $z \geq 1$ and up to $\ell\sim500$ at $z$=0.1.

\subsection{Halo field}
\label{sec:halo}

\begin{figure*}
    \centering
    \includegraphics[width=\textwidth]{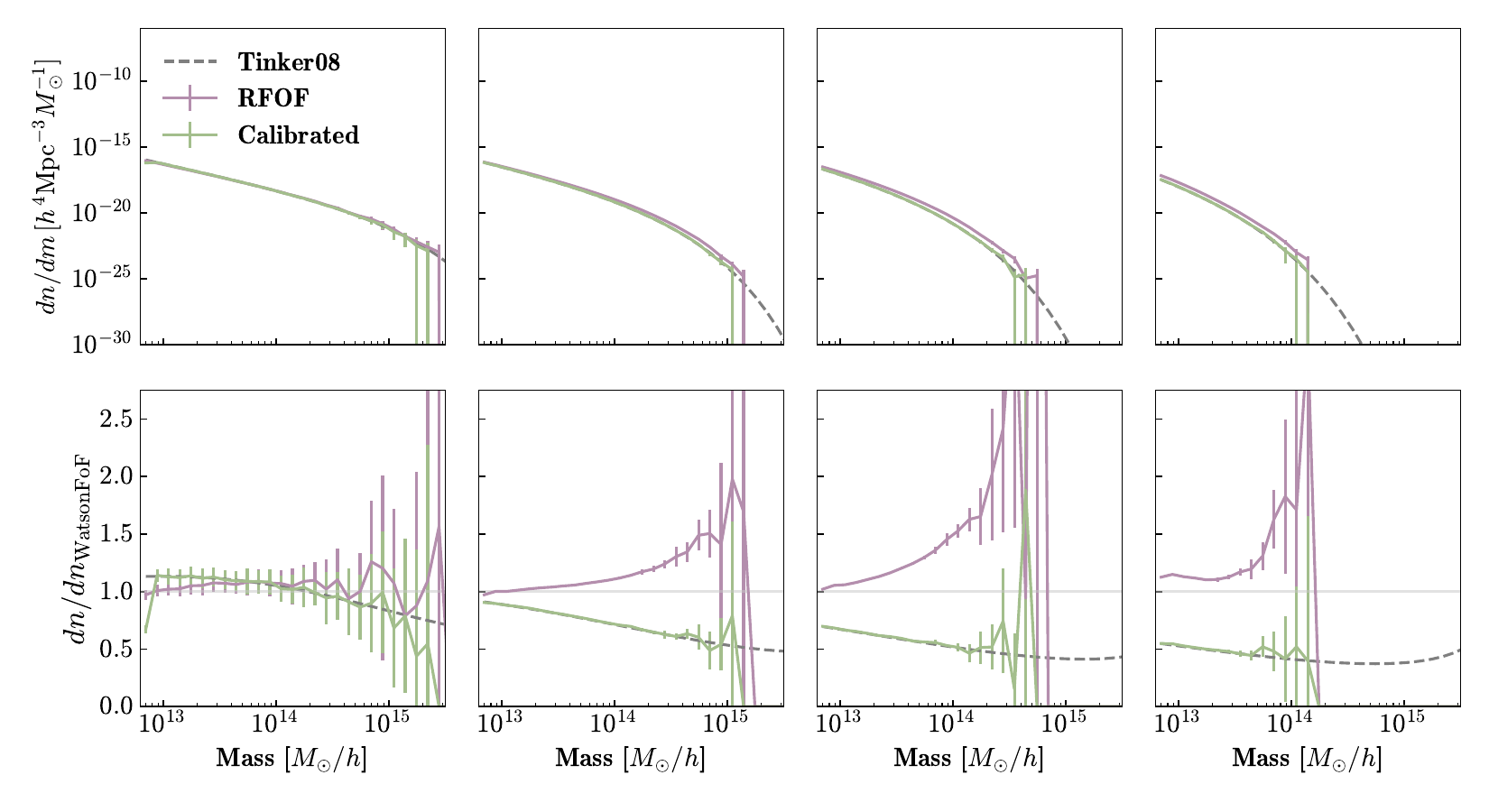}
    
    \caption{ 
    {\bf Top panels}: halo mass functions of HalfDome simulations of different resolutions and of Tinker08 fitting function~\cite{Tinker2008}, at $z$=0, 1, 2, 3. {\bf Bottom panels}: the ratios between HalfDome runs and a fitting function for FoF halos \citep{watson2013}. Masses calibrated to Tinker08 are shown in green. Error bars are the standard deviation of 11 realizations.
    It can be seen that calibrated masses agree with Tinker08. 
    }
    \label{fig:HMF}
\end{figure*}

Fig.~\ref{fig:HMF} shows the halo mass functions (HMF) computed from our simulations at full resolution, in comparison to fitting functions for Friends-of-Friends (FoF) \cite{watson2013} and $M_{200m}$ from Tinker08. The HMF $dn/dm$ is defined as the comoving number density of halos per unit halo mass. We divide the halos into bins spaced evenly by $\Delta (\log_{10} M_h/M_\odot) = 0.1$. 
In the top panel we show the HMF from RFoF masses, as well as for the same halos with masses calibrated to Tinker08. 
In the bottom panel we show the ratios of HMFs from our simulations relative to the Watson-FoF fitting function \cite{watson2013}. We also show the fitting function from Tinker08 with the same normalization, demonstrating agreement between the target mass definition and our recalibration scheme. We show results at $z$=0, 1, 2, 3.

\begin{figure*}
    \centering
    \includegraphics[width=1.0\textwidth]{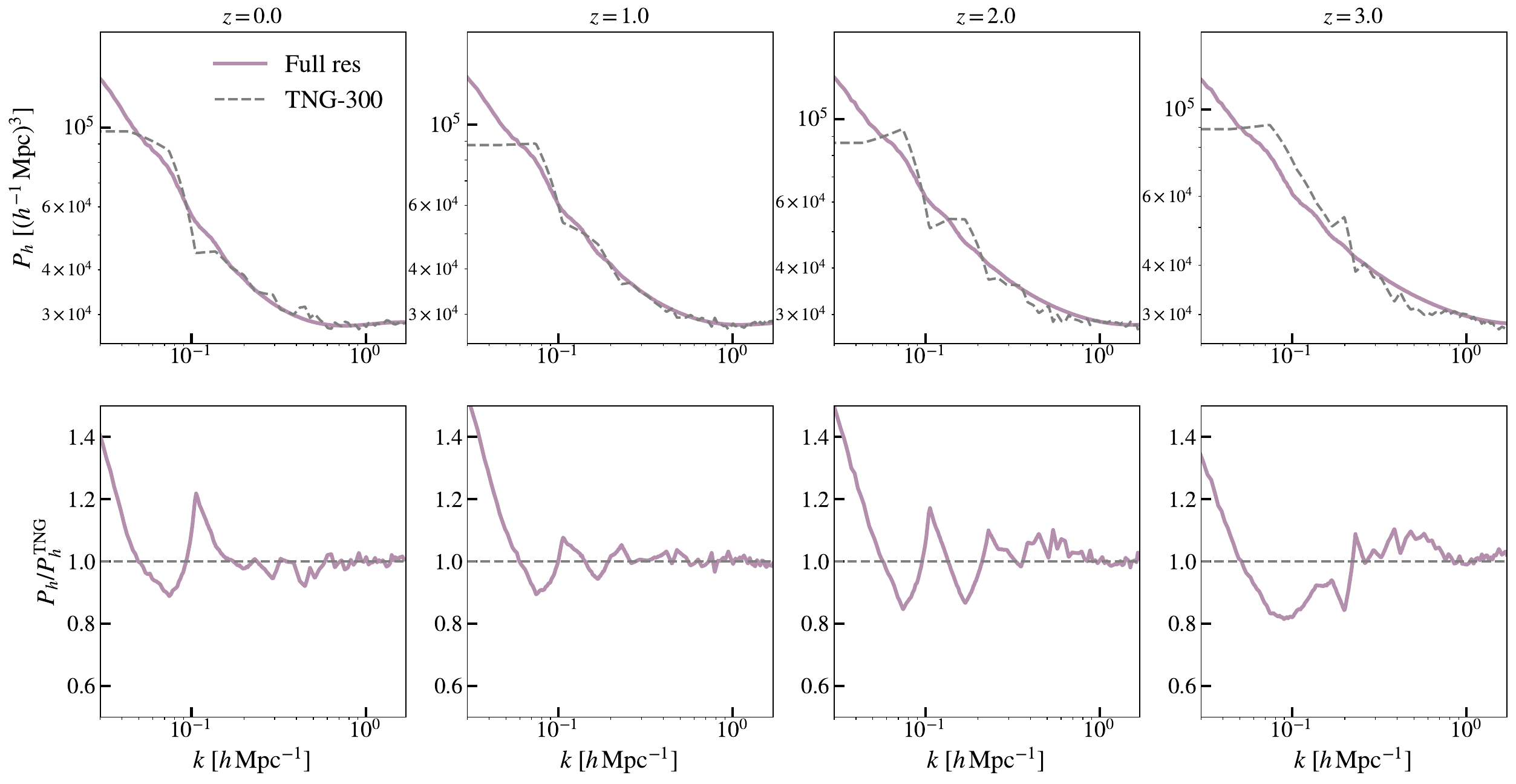}
    \caption{
    {\bf Top panels}: halo power spectra of full resolution HalfDome, averaged over all 11 simulations, and TNG300-Dark-1, at $z$=0, 1, 2, 3. We applied a mass cut of $10^{14}$~\msunh at $z$=0 and used abundance matching to select halos at higher redshifts. {\bf Bottom panels}: the ratios between the HalfDome runs and TNG. The large variance in the
    bottom panels is the result of cosmic variance from the single small TNG box.}
    \label{fig:Phh}
\end{figure*}

Fig.~\ref{fig:Phh} compares the mean halo power spectra of our full resolution runs with that of TNG300-1-Dark~\cite{2019ComAC...6....2N}, which is a dark-matter-only simulation in a box with side-length 205~\mpch and $2500^3$ dark matter particles. TNG uses FoF with a linking length of 0.2 to find halos. 
To select the halos, we apply a mass cut of $10^{14}$~\msunh at $z$=0 and use abundance matching to select halos at higher redshifts, i.e.~we select the most massive halos at each redshift such that the number of halos matches the number that passed the mass cut at $z$=0, for HalfDome and TNG respectively. This leads to a minimum mass cut of $\sim6\times10^{12}$~\msunh at $z=3$. The overall agreement is of order a few percent, with some discrepancy due to noise in the TNG power spectrum due to it being a single relatively small box. 

\subsection{Primordial non-Gaussianity}
\label{sec:fnl}

\begin{figure*}
    \centering
    \includegraphics[width=1\linewidth]{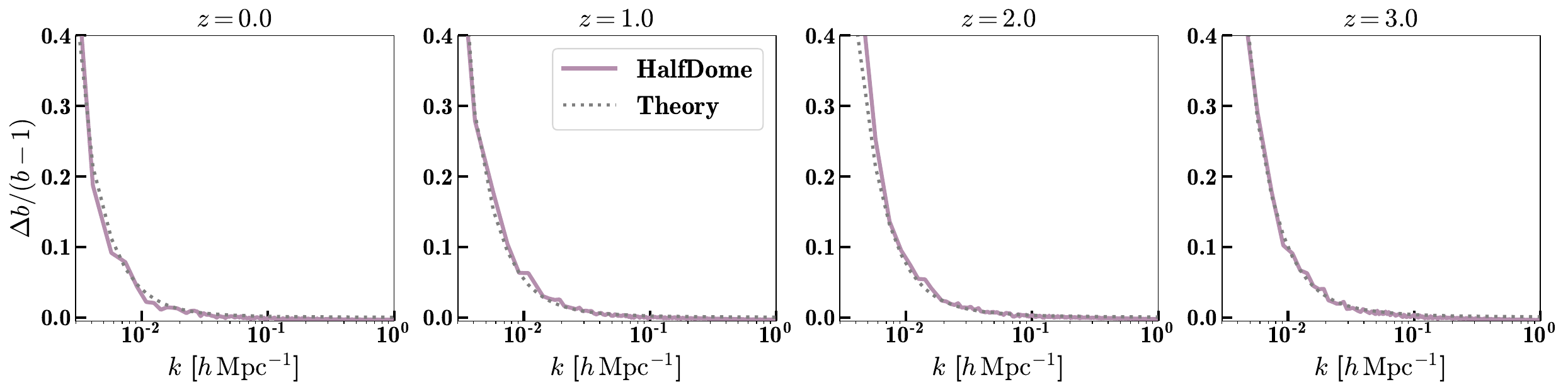}
    \caption{$\Delta b / (b-1)$ computed from HalfDome full-resolution simulation, where $b$ is the halo bias and $\Delta b$ is the difference between $f_{\rm NL}$=20 and $f_{\rm NL}$=0 runs (seed matched). The theoretical prediction is based on Eq.~\ref{fnl_theory}.
    } 
    \label{fig:fnl_comparison}
\end{figure*}

Measuring primordial non-Gaussianity, quantified as $f_{\rm NL}$, is a major scientific goal for upcoming cosmological surveys~\cite{{2001PhRvD..63f3002K,2009JCAP...09..006S,2009ApJ...703.1230J}}. Here we focus on local-type primordial non-Gaussianity, which modifies the gravitational potential as follows, 
\begin{align}
\Phi=\phi+f_{\mathrm{NL}} ( \phi^2 - \langle \phi^2 \rangle),
\end{align}
where $\phi$ is the primordial potential assumed to be a Gaussian random field, and $f_{\rm NL}$ is the amplitude of non-Gaussianity. While single-field inflation models predict $f_{\rm NL} = 0$, many multi-field models predict non-zero $f_{\rm NL}$. Measuring a non-zero $f_{\rm NL}$ would be a major step toward understanding inflation. Upcoming cosmological surveys are expected to be sensitive to $f_{\rm NL}$ of order unity~\citep[e.g.,][]{Planck:2019kim,Mueller:2021jvn,DESI:2023duv}.

Local-type primoridal non-Gaussianity leaves a scale dependent halo bias,
\begin{align} \label{halo_bias}
b(k)=\sqrt{\frac{P_{\mathrm{h}}(k)}{P_{\mathrm{m}}(k)}}
\end{align}
where $P_{\mathrm{h}}$ and $P_{\mathrm{m}}$ are the halo power spectrum and matter power spectrum, respectively. It has been shown that the difference in bias between a cosmology with non-zero $f_{\rm NL}$ and $f_{\rm NL}=0$ is given by
\begin{align}
\label{fnl_theory}
\Delta b(M, k)=3 f_{\mathrm{NL}}(b-1) \delta_c \frac{\Omega_m H^2_0}{k^2 T(k) D(z)},
\end{align}
where $\delta_c \sim 1.68$ is the present-day critical density threshold, $T(k)$ the matter transfer function, $D(z)$ the growth factor normalized to the scale factor $a$ in the matter era, and $H_0$ the Hubble parameter \citep{2008PhRvD..77l3514D,2008ApJ...677L..77M,2009JCAP...03..004S,2009MNRAS.396...85D}.

As part of our initial data release we include a simulation with local-type non-Gaussianity with $f_{\mathrm{NL}}$=20, with all other parameters the same as the full-resolution run (Table~\ref{Tab:sims}) and matched to one of the random seeds. Fig.~\ref{fig:fnl_comparison} compares the scale-dependent bias observed in our simulations, computed using Eq.~\ref{halo_bias} with the same abundance matching as in Sec.~\ref{sec:halo}, compared with the theoretical prediciton of Eq.~\ref{fnl_theory}, showing excellent agreement.

\section{Conclusions}
\label{sec:conclusion}

In this paper, we present the initial data release of the HalfDome cosmological simulations, including lightcones and halo catalogs, designed specifically for the joint analysis of Stage-IV cosmological surveys. Our full-resolution runs ($6144^3$ particles, 3.75~\gpch box) include 11 realizations at a fixed cosmology ---  to enable uncertainty quantification in survey analysis pipeline testing --- and an additional run with non-zero primordial non-Gaussianity ($f_{\rm NL}$=20). Our simulations reach a minimum halo mass of $\sim 6 \times 10^{12}$~\msunh and resolution $k_{\rm max}$$\sim$1~\hmpc. 
Our data products include (1) downsampled particles, (2) halo catalogs, (3) mass sheets, and (4) velocity sheets, all between redshift $z$=0--4 (Sec.~\ref{sec:method}). The detailed simulation setup and runtime information can be found in Table~\ref{Tab:sims}. We conducted careful convergence tests and validated our results against higher-resolution simulations and theoretical predictions  (Sec.~\ref{sec:result}). 


We envisage building a coherent and collaborative ecosystem for the joint analysis of correlated astronomical data, where different components in the pipeline --- such as the N-body simulation or the model of CMB secondaries --- can be flexibly added and substituted depending on the application. Our initial N-body data release presented in this paper provides the backbone for the development of such an ecosystem. Our forthcoming papers will present realistic mocks that model multi-wavelength observables, applicable to LSS and CMB surveys, such as LSST, Euclid, SPHEREx, Roman, DESI, PFS, Simons Observatory, CMB-S4, and LiteBIRD. We are also actively interested in collaborating to develop mocks for other observables such as line intensity mapping, Lyman-$\alpha$ forest, X-ray, Fast Radio Bursts, and beyond. 
These simulations will enable the study of a wide range of topics, such as galaxies, clusters, weak lensing, CMB secondaries, cross-correlations, non-Gaussian statistics, baryonic effects, and more. In addition to being a tool for traditional cosmological analysis, we aim to pave a path to simulation-based inference and the application of other machine learning methods. 

Our data is publicly available:
instructions to access the data and plans for future data releases can be found at \url{https://halfdomesims.github.io}.

\begin{acknowledgments}
We thank 
Marcelo Alvarez, 
Linda Blot, 
Will Coulton, 
Biwei Dai,
Adriaan Duivenvoorden,
Jo Dunkley,
Colin Hill, 
Elisabeth Krause,
Alex Laguë, 
Mathew Madhavacheril,
Yuuki Omori,
Giuseppe Puglisi,
Neelima Sehgal,
Uroš Seljak,
David Spergel,
James Sullivan,
Masahiro Takada, 
Hideki Tanimura, 
Leander Thiele, 
Beatriz Tucci,
Francisco Villaescusa-Navaro, 
and Junjie Xia
for insightful discussion. 
We thank Julian Borrill and Reijo Keskitalo for their support with computational resources and data storage. We thank James Sunseri for designing the HalfDome logo. 
This work was supported by JSPS KAKENHI Grants JP23KJ0392 (YZ), 23K13095 and 23H00107 (JL). 
This research used resources of the National Energy Research Scientific Computing Center (NERSC), a U.S.~Department of Energy Office of Science User Facility located at Lawrence Berkeley National Laboratory, operated under Contract No.~DE-AC02-05CH11231 using NERSC award  HEP-ERCAP0023125. The authors acknowledge the Texas Advanced Computing Center (TACC) at The University of Texas at Austin for providing grid resources that have contributed to the research results reported within this paper. This work was supported by MEXT as ``Program for Promoting Researches on the Supercomputer Fugaku'' (Multi-wavelength Cosmological Simulations for Next-generation Surveys, JPMXP1020230407) and used computational resources of supercomputer Fugaku provided by the RIKEN Center for Computational Science (Project ID: hp230202). We use \texttt{FastPM}~\cite{Feng2016, Bayer_2021_fastpm} to simulate large-scale structure. 
We use \texttt{nbodykit}~\cite{Hand_2018} to compute overdensity fields and power spectra of the simulations. We use \texttt{Class} to compute the theoretical power spectra~\cite{classy}. 

\end{acknowledgments}

\bibliographystyle{physrev}
\bibliography{references}

\end{document}